\documentclass[draft,numberedheadings]{aipproc}

\layoutstyle{8x11single}

\newcommand{\be}{\begin{equation}}
\newcommand{\ee}{\end{equation}}
\newcommand{\ba}{\begin{eqnarray}}
\newcommand{\ea}{\end{eqnarray}}
\newcommand{\onehalf}{\textstyle{\frac{1}{2}}}
\def\qed{\hbox{${\vcenter{\vbox{
   \hrule height 0.4pt\hbox{\vrule width 0.4pt height 6pt
   \kern5pt\vrule width 0.4pt}\hrule height 0.4pt}}}$}}
\begin{document}
\title[Cosmological Constant and Fundamental Physics]{Some Implications of the 
Cosmological Constant\\ to Fundamental Physics}



\author{R. Aldrovandi}{
  address={Instituto de F\'{\i}sica Te\'orica,
Universidade Estadual Paulista \\
Rua Pamplona 145, 01405-900 S\~ao Paulo, Brazil}}

\author{J. P. Beltr\'an Almeida}{
  address={Instituto de F\'{\i}sica Te\'orica,
Universidade Estadual Paulista \\
Rua Pamplona 145, 01405-900 S\~ao Paulo, Brazil}}

\author{J. G. Pereira}{
  address={Instituto de F\'{\i}sica Te\'orica,
Universidade Estadual Paulista \\
Rua Pamplona 145, 01405-900 S\~ao Paulo, Brazil}}

\begin{abstract}
In the presence of a cosmological constant, ordinary Poincar\'e special relativity 
is no longer valid and must be replaced by a de Sitter special relativity, in 
which Minkowski space is replaced by  a de Sitter spacetime. In consequence, the 
ordinary notions of energy and momentum change, and will satisfy a different 
kinematic relation. Such a theory is a different kind of a doubly special 
relativity. Since the only difference between the Poincar\'e and the de Sitter 
groups is the replacement of translations by certain linear combinations of 
translations and proper conformal transformations, the net result of this change 
is ultimately the breakdown of ordinary translational invariance. From the 
experimental point of view, therefore, a de Sitter special relativity might be 
probed by looking for possible violations of translational invariance. If we 
assume the existence of a connection between the energy scale of an experiment and 
the local value of the cosmological constant, there would be changes in the 
kinematics of massive particles which could hopefully be detected in high-energy 
experiments. Furthermore, due to the presence of a horizon, the usual causal 
structure of spacetime would be significantly modified at the Planck scale.
\end{abstract}
\maketitle

\section{Introduction}

When the cosmological constant $\Lambda$ vanishes, absence of gravitation is 
represented by Minkowski spacetime, a solution of the sourceless Einstein's 
equation. Its isometry transformations are those of the Poincar\'e group, which is 
the group governing the kinematics of special relativity. For a non-vanishing 
$\Lambda$, however, Minkowski is no longer a solution of the corresponding 
Einstein's equation and becomes, in this sense, physically meaningless. In this 
case, absence of gravitation turns out to be represented by the de Sitter 
spacetime. Now, the group governing the kinematics in a de Sitter spacetime is not 
the Poincar\'e, but the de Sitter group. This means essentially that, in the 
presence of a non-vanishing $\Lambda$, ordinary Poincar\'e special relativity will 
no longer be valid, and must be replaced by a de Sitter special relativity 
\cite{dssr,guoatall}. Since the {\it local} symmetry is also represented by the de 
Sitter group, the tangent space at each point of a Riemannian spacetime must also 
be replaced by an osculating de Sitter space.

An important point of this construction is that it retains the quotient character 
of spacetime and, consequently, a notion of homogeneity. As in ordinary special 
relativity, therefore, whose underlying  Minkowski spacetime is the quotient space 
of the Poincar\'e by the Lorentz groups, the underlying spacetime of the de Sitter 
relativity will be the quotient space of the de Sitter and the Lorentz groups. 
Now, a space is said to be transitive under a set of transformations --- or 
homogeneous under them --- when any two points of it can be attained from each 
other by a transformation belonging to the set. For example, the Minkowski 
spacetime is transitive under spacetime translations. However, the de Sitter 
spacetime is transitive, not under translations, but under a combination of 
translations and proper conformal transformations, with the relative importance of 
these contributions being determined by the value of the cosmological constant. 
Observe that, due to the quotient character of spacetime, it will respond 
concomitantly to any deformation occurring in the symmetry groups. For small 
values of $\Lambda$, for example, the de Sitter group will approach the Poincar\'e 
group, and the underlying spacetime will approach the flat Minkowski spacetime. 
For large values of $\Lambda$, the de Sitter group will approach the conformal 
Poincar\'e group, and the underlying spacetime will approach a new maximally-
symmetric conic spacetime, which is homogeneous under proper conformal 
transformations \cite{confor}.

An immediate consequence of the de Sitter homogeneity properties is that the 
ordinary notions of energy and momentum will change \cite{aap}. In fact, the 
conserved momentum, for example, will now be obtained from the invariance of the 
physical system, not under translations, but under a combination of translations 
and proper conformal transformations. The conserved momentum, therefore, will be 
given by a combination of ordinary and proper conformal momenta, the relative 
importance of these contributions being again determined by the value of the 
cosmological constant. Of course, the usual special relativistic relation between 
ordinary mass, energy and momentum will also change. Another important consequence 
of the presence of a cosmological constant is that it modifies the usual Lorentz 
causal structure of spacetime, defined by the light cone. In fact, the causal 
domain of all observers will be further restricted by the presence of the de 
Sitter horizon.

To get some insight on how the de Sitter special relativity may be thought of, let 
us briefly recall the relationship between the de Sitter and the Galilei groups, 
which comes from the Wigner--In\"on\"u processes of group contraction and 
expansion \cite{inonu,gil}. Ordinary Poincar\'e special relativity can be viewed 
as describing the implications to Galilei's relativity of introducing a 
fundamental velocity-scale into the Galilei group. Conversely, the latter can be 
obtained from the special-relativistic Poincar\'e group by taking the formal limit 
of the velocity scale going to infinity (non-relativistic limit). We can, in an 
analogous way, say that de Sitter relativity describes the implications to 
Galilei's relativity of introducing both a velocity and a length scales in the 
Galilei group. In the formal limit of the length-scale going to infinity, the de 
Sitter groups contract to the Poincar\'e group, in which only the velocity scale 
is present. It is interesting to observe that the order of the group expansions 
(or contractions) is not important. If we introduce in the Galilei group a 
fundamental length parameter, we end up with the Newton-Hooke group \cite{nh}, 
which describes a (non-relativistic) relativity in the presence of a cosmological 
constant \cite{gibb}. Adding to this group a fundamental velocity scale, we end up 
again with the de Sitter group, whose underlying relativity involves both a 
velocity and a length scales. Conversely, the low-velocity limit of the de Sitter 
group yields the Newton-Hooke group, which contracts to the Galilei group in the 
limit of a vanishing cosmological constant.

The purpose of this article is to discuss the main features and consequences of a 
de Sitter special relativity. We will proceed as follows. Section~2 is a review of 
the fundamental properties of the de Sitter groups and spaces. In section~3 it is 
shown how the transitivity properties of the de Sitter spacetime can lead to two 
different notions of distance: a translational and a conformal distance. In 
section 4, the fundamentals of a de Sitter special relativity are presented and 
discussed. In particular, an analysis of the deformed generators acting on the de 
Sitter space is made, which allows us to understand how a de Sitter relativity 
keeps a well defined algebraic structure on a de Sitter spacetime. The modified 
notions of energy and momentum, as well as the new relationship between them, are 
obtained in section 5. Finally, section 6 discusses the results obtained. In 
particular, the relation of the de Sitter relativity with the so called doubly 
special relativity is discussed, and some speculations on possible 
phenomenological consequences are examined. Closing the article, an analysis of 
the new spacetime causal structure at the Planck length is performed.

\section{de Sitter spacetimes and groups}

Spacetimes with constant scalar curvature $R$ are maximally symmetric: they can 
lodge the highest possible number of Killing vectors. Given a metric signature, 
this spacetime is unique~\cite{weinberg} for each value of $R$. Minkowski 
spacetime $M$, with $R = 0$, is the simplest one. Its group of motions is the 
Poin\-ca\-r\'e group ${\mathcal P} = {\mathcal L} \oslash {\mathcal T}$, the semi-
direct product of the Lorentz ${\mathcal L} = SO(3,1)$ and the translation group 
${\mathcal T}$. The latter acts transitively on $M$ and its group manifold can be 
identified with $M$. Indeed,  Minkowski  spacetime is a homogeneous space under 
${\mathcal P}$, actually the quotient
\[
M = {\mathcal P}/{\mathcal L}.
\]

Amongst curved spacetimes, the de Sitter and anti-de Sitter spaces are the only 
possibilities \cite{ellis}. One of them has negative, and the other has positive 
scalar curvature. They are hyper-surfaces in the ``host'' pseudo-Euclidean spaces 
${\bf E}^{4,1}$ and ${\bf E}^{3,2}$, inclusions whose points in Cartesian 
coordinates $(\chi^A) = (\chi^0,
\chi^1, \chi^2, \chi^3, \chi^{4})$ satisfy respectively
\[
\eta_{AB} \chi^A \chi^B \equiv (\chi^0)^2 - (\chi^1)^2 -
(\chi^2)^2 - (\chi^3)^2 - (\chi^{4})^2 = -\, l^2
\]
and
\[
\eta_{AB} \chi^A \chi^B \equiv (\chi^0)^2 - (\chi^1)^2 -
(\chi^2)^2 - (\chi^3)^2 + (\chi^{4})^2 = l^2,
\]
where $l$ is the so-called de Sitter length-parameter. We use the Latin alphabet 
($a, b, c \dots = 0,1,2,3$) to denote the four-dimensional algebra and tangent 
space indices, whose metric tensor is $\eta_{a b} = $ diag $(1$, $-1$, $-1$, $-
1)$. Defining the dimensionless coordinate $\chi^{\prime 4} = \chi^4/l$, and using 
the notation ${\sf s} = \eta_{44}$, the above conditions can be put together as
\be
\frac{1}{l^2} \, \eta_{a b} \, \chi^{a} \chi^{b} + {\sf s} \, (\chi^{\prime 4})^2 
= {\sf s}.
\label{dspace1}
\ee
For ${\sf s} = - 1$, we have the de Sitter space $d S = dS(4,1)$, whose metric is 
induced from the pseudo-Euclidean metric $\eta_{AB}$ = $(+1,-1,-1,-1,-1)$. It has 
the pseudo-orthogonal group $SO(4,1)$ as group of motions. Sign ${\sf s} = + 1$ 
corresponds to  anti-de Sitter space, denoted by $AdS = dS(3,2)$. It comes from 
$\eta_{AB}$ = $(+1,-1,-1,-1,+1)$, and has $SO(3,2)$ as its group of motions. Both 
spaces are homogeneous~\cite{livro}:
\[
dS = dS(4,1) = SO(4,1)/ {\mathcal L} \quad {\rm and} \quad AdS = dS(3,2) = 
SO(3,2)/ {\mathcal L}.
\]
Furthermore, they are solutions of the sourceless Einstein's equation, provided 
the cosmological constant $\Lambda$ and the de Sitter length-parameter $l$ are 
related by
\be
\Lambda = -\, \frac{3 {\sf s}}{l^2} \,\, .
\label{lambdaR}
\ee

\subsection{Stereographic coordinates}

To comply with the observational data \cite{obs}, we consider from now on the $dS$ 
case, for which $\Lambda \ge 0$ and
\be
-\,  \frac{1}{l^2} \, \eta_{a b} \, \chi^{a} \chi^{b} + (\chi^{\prime 4})^2 = 1.
\label{dspace1b}
\ee
The four-dimensional stereographic coordinates $\{x^a\}$ are obtained through a 
projection from the de Sitter hyper-surface into a target Minkowski spacetime. 
They are given by \cite{gursey}
\be
\chi^{a} = \Omega(x) \, x^a \quad \mbox{and} \quad
\chi'^4 = -\, \Omega(x) \left(1 + \frac{\sigma^2}{4 l^2} \right),
\label{xi4}
\ee 
where
\be
\Omega(x) = \left(1 - \frac{\sigma^2}{4 l^2}\right)^{-1},
\label{n}
\ee
with $\sigma^2 = \eta_{a b} \, x^a x^b$. The $\{x^a\}$ take values on the  
Minkowski spacetime on which the stereographic projection is made.

\subsection{Kinematic groups: transitivity}

In terms of the host-space  Cartesian coordinates $\chi^A$, the generators of the 
infinitesimal de Sitter transformations are
\be
L_{A B} = \eta_{AC} \, \chi^C \, \frac{\partial}{\partial \chi^B} -
\eta_{BC} \, \chi^C \, \frac{\partial}{\partial \chi^A} \, \, ,
\label{dsgene}
\ee
which satisfy the commutation relations
\be
\left[ L_{AB}, L_{CD} \right] = \eta_{BC} L_{AD} + \eta_{AD} L_{BC} - \eta_{BD} 
L_{AC}
- \eta_{AC} L_{BD} .
\label{desla}
\ee
In terms of the stereographic coordinates $\{x^a\}$, these generators are written 
as
\be
L_{ab} =
\eta_{ac} \, x^c \, P_b - \eta_{bc} \, x^c \, P_a
\label{cp0}
\ee
and
\be
L_{a4} = l P_a - ({4 l})^{-1} K_a,
\label{dstra}
\ee
where
\be
P_a = {\partial}/{\partial x^a}
\ee
are the translation generators (with dimension of {\it length}$^{-1}$), and
\be
K_a = \left(2 \eta_{ab} \, x^b x^c - \sigma^2 \delta_{a}{}^{c} \right) P_c
\ee
are the generators of {\it proper} conformal transformations (with dimension of 
{\it length}). 
The generators $L_{a b}$ refer to the Lorentz subgroup of de Sitter, whereas 
$L_{a4}$ define transitivity on the corresponding de Sitter space. For this 
reason, they are usually called the de Sitter ``translation'' generators. The 
crucial point is to observe, as implied by Eq.~(\ref{dstra}), the de Sitter 
spacetime is transitive under a combination of translations and proper conformal 
transformations. The relative importance of each one of these transformations is 
determined by the value of the length parameter $l$, that is, by the value of the 
cosmological constant. 

\subsection{Contraction limits}

The group contraction procedure requires that, before each limit is taken,  the 
generators be modified through an appropriate insertion of parameters.  These 
alterations are frequently guided by dimensional considerations and are different 
for different limits \cite{inonu2}. For this reason, the 
$\Lambda \rightarrow  0\, $ and the $\Lambda \rightarrow  \infty\,\,  $  limits 
must be considered separately.

\subsubsection{Vanishing cosmological constant limit}

To study the limit of a vanishing cosmological constant ($l \to \infty$), we 
rewrite the de Sitter generators in the forms
\be
L_{ab} =
\eta_{ac} \, x^c \, P_b - \eta_{bc} \, x^c \, P_a
\label{dslore}
\ee
and
\be
\Pi_a \equiv \frac{L_{a4}}{l} =
P_a - \frac{1}{4 l^2}\,  K_a\, .
\label{l0}
\ee
The  $L_{ab}$'s generate the Lorentz transformations on Minkowski spacetime, and 
satisfy the commutation relation
\be
\left[ L_{ab}, L_{cd} \right] = \eta_{bc} L_{ad} + \eta_{ad} L_{bc} - \eta_{bd} 
L_{ac}
- \eta_{ac} L_{bd}.
\ee
For $l \to \infty$, the generators $\Pi_a$ reduce to ordinary translations, and 
the de Sitter group contracts to the Poincar\'e group ${\mathcal P} = {\mathcal L} 
\oslash {\mathcal T}$. Concomitant with the algebra and group deformations, the de 
Sitter space $dS(4,1)$ reduces to the Minkowski spacetime $M = {\mathcal 
P}/{\mathcal L}$, which is transitive under ordinary translations only.

\subsubsection{Infinite cosmological constant limit}

We remark, to begin with, that the limit $\Lambda \to \infty\,\,$ has to be 
understood as purely formal. In fact, considering that it corresponds to the small 
distance limit $l \to 0$, quantum effects should necessarily be taken into 
account. Such effects, as is well known, provide  for $\Lambda$ a cutoff value 
which prevents the limit to be physically achieved.  Actually, the reference value 
for defining small and large $\Lambda$ is the Planck cosmological constant 
$\Lambda_P = 3/l_P^2$, with $l_P$ the Planck length. A small $\Lambda$ will then 
be characterized by $\Lambda \, l_P^2 \to 0$, and a large $\Lambda$  by $\Lambda 
\, l_P^2 \to 1$. 

The interest of  the  $\Lambda \to \infty\,\,$ limit lies in the fact that it 
yields the algebraic and geometric structures behind the  ``classical'' physics of 
extremely high $\Lambda$. 
Let us then recall that the de Sitter space tends, in this case, to a cone 
spacetime, denoted $N$, which is related to Minkowski $M$ through the spacetime 
inversion \cite{confor}
\be
x^a \to -\,  \frac{x^a}{\sigma^2} \,\, .
\label{inversion}
\ee
By this inversion, the points at infinity on $M$ are led into the vertex of the 
cone-space $N$, and those on the light-cone of $M$ become the infinity of $N$. 
Applying  it to the Minkowski interval
\be
ds^2 = \eta_{ab} \, dx^a dx^b,
\ee
we see that\footnote{In addition to denoting the indices of the Minkowski 
spacetime $M$, the Latin alphabet ($a, b, c \dots = 0,1,2,3$) will also be used to 
denote the algebra and space indices of the cone spacetime $N$.}
\be
ds^2 \to d\bar{s}^2 = {\bar{\eta}}_{ab} \, d{x}^a d{x}^b,
\label{confin}
\ee
where
\be
{\bar{\eta}}_{ab} = {\sigma}^{-4} \, \eta_{ab}, \qquad {\bar{\eta}}^{ab} =
{\sigma}^{4} \, \eta^{ab}
\label{Nmetric}
\ee
is the metric on the cone-space $N$. It is important to recall also that the 
spacetime inversion (\ref{inversion}) is well known to relate translations with 
proper conformal transformations \cite{coleman}:
\be
P_a \rightarrow K_a.
\ee
On the other hand, it is found that the Lorentz generators do not  change:
\be
L_{ab} \to L_{ab}.
\ee

The above results imply that, to study the limit of an infinite cosmological 
constant ($l \to 0$), it is necessary to write the de Sitter generators in the 
form
\be
~~~~~~~\bar{L}_{ab} \equiv \sigma^{-4} L_{ab} =
\bar{\eta}_{ac} \, x^c \, P_b - \bar{\eta}_{bc} \, x^c \, P_a
\label{dSLgbis}
\ee
and
\be
\bar{\Pi}_a \equiv 4 l \, L_{a4} = 4 l^2 P_a - K_a.~~~~~~~~
\label{linf}
\ee
The generators $\bar{L}_{ab}$ satisfy the commutation relation
\be
\left[ \bar{L}_{ab}, \bar{L}_{cd} \right] = \bar{\eta}_{bc} \bar{L}_{ad} + 
\bar{\eta}_{ad} \bar{L}_{bc} - \bar{\eta}_{bd} \bar{L}_{ac} - \bar{\eta}_{ac} 
\bar{L}_{bd}.
\ee
Since $\bar{L}_{ab}$ satisfy Lorentz-like commutations relations, they can be 
interpreted as the generators of {\it conformal Lorentz transformations}, whose 
group we denote by $\bar{\mathcal L}$. For $l \to 0$, the generators $\bar{\Pi}_a$ 
reduce to (minus) the proper conformal generators, whose group we denote by 
$\bar{\mathcal T}$. In this limit, therefore, the de Sitter group contracts to the 
{\em conformal}\, Poincar\'e group $\bar{\mathcal P} = \bar{\mathcal L} \oslash 
\bar{\mathcal T}$, the semi-direct product of the conformal Lorentz and the proper 
conformal groups \cite{ap1}. As can easily be verified, the interval $d\bar{s}^2$ 
is invariant under $\bar{\mathcal P}$. Concomitant with the group contraction $l 
\to 0$, the de Sitter space reduces to the conic spacetime
\[
N = \bar{\mathcal P}/\bar{\mathcal L},
\]
which is transitive under proper conformal transformations.

\section{Transitivity and the notion of distance}

The two concurrent, but different types of transformations present in the 
generators defining transitivity on the de Sitter spacetime $dS$ give rise to two 
different notions of distance: one which is related to translations, and another 
which  is related to proper conformal transformations. This means that it is 
possible to define two different metrics on  $dS$, one invariant under 
translations, and another invariant under proper conformal transformations. As a 
consequence, there will be two different family of geodesics, one joining all 
points equivalent under translations, and another joining all points equivalent 
under proper conformal transformations. If one considers only one of these 
families, therefore, there will be points on  $dS$ which cannot be joined to each 
other by any geodesic. This is a well known property of the de Sitter spacetime 
\cite{ellis}. In what follows we explore these concepts in more detail.

The Greek alphabet ($\mu, \nu, \rho, \dots =0,1,2,3$) will be used to denote 
indices related to the de Sitter spacetime. For example, its coordinates will be 
denoted by $\{x^\mu\}$. We recall that the Latin alphabet ($a, b, c \dots = 
0,1,2,3$) relate to the de Sitter algebra, as well as to the spacetime indices of 
both limits of the $dS$ spacetime: Minkowski space ($\Lambda \rightarrow  0\, $  
limit) and the cone spacetime ($\Lambda \rightarrow  \infty\,\,  $ limit). This 
allows the introduction of the hol\-o\-nomic tetrad $\delta^{a}{}_{\mu}$, which 
satisfies
\be
\eta_{\mu \nu} = \delta^{a}{}_{\mu} \delta^{b}{}_{\nu} \, \eta_{ab}, \quad
\bar{\eta}_{\mu \nu} = \delta^{a}{}_{\mu} \delta^{b}{}_{\nu} \, \bar{\eta}_{ab}.
\ee
Consequently, we can also write
\be
\sigma^2 = \eta_{ab} \, x^a x^b = \eta_{\mu \nu} \, x^\mu x^\nu, \quad
\bar{\sigma}^2 = \bar{\eta}_{ab} \, x^a x^b = \bar{\eta}_{\mu \nu} \, x^\mu x^\nu,
\ee
where we have identified $x^a = \delta^{a}{}_{\mu} x^\mu$. 

\subsection{Translational distance}

The first notion of distance is that related to translations, which  become the 
dominant part of the $dS$ transitivity generators for small values of $\Lambda$. 
To study its properties it is, therefore, necessary to use a parameterization 
appropriate for the limit $\Lambda \to 0$.  This parameterization is naturally 
provided by Eq.~(\ref{dspace1}),
\be
K_G \, {\Omega}^2(x) \, {\sigma}^2 + (\chi'^4)^2 = 1,
\ee
where the Gaussian curvature 
\be
K_G = -\, 1/l^2
\ee
of  the $dS$ spacetime turns up. We introduce now the an\-hol\-o\-no\-mic tetrad 
field
\be
h^a{}_\mu = \Omega \, \delta^a{}_\mu.
\label{tetraze}
\ee
If $\eta_{ab}$ denotes the Minkowski metric, the de Sitter metric can, in this 
case, be written as
\be
g_{\mu \nu}  \equiv   h^{a}{}_{\mu} \, h^{b}{}_{\nu} \, \eta_{a b} =
\Omega^2(x) \, \eta_{\mu \nu}.
\label{44}
\ee
It defines the ``translational distance'', with squared interval 
\be
d\tau^2 = g_{\mu \nu} \, dx^\mu dx^\nu \equiv
\Omega^2(x) \, \eta_{\mu \nu} \, dx^\mu dx^\nu.
\label{onod}
\ee
For $l \to \infty$ ($\Lambda \to 0$), $dS$ contracts to the Minkowski spacetime 
$M$, and $d\tau^2$ reduces to the Lorentz-invariant interval
\be
d\tau^2 \to ds^2 = \eta_{\mu \nu} \, dx^\mu dx^\nu.
\ee
Due to its translational transitivity, this is the only notion of distance that 
can be defined on $M$.

\subsection{Conformal distance}

The second notion of distance is that related to the proper conformal 
transformation. Since this transformation is the most important part of the 
transitivity generators for large values of $\Lambda$, its study requires a 
parameterization appropriate for the limit $\Lambda \to \infty$. This can be 
achieved by rewriting Eq.~(\ref{dspace1}) in the form
\be
\bar{K}_G\, \bar{\Omega}^2(x) \, \bar{\sigma}^{2} + (\chi'^4)^2 = 1,
\ee
where
\be
\bar{\Omega}(x) \equiv \frac{\sigma^2}{4 l^2} \, \Omega(x) = -\, 
\frac{1}{(1 - {4 l^2}/{\sigma^2} )}
\label{nbar}
\ee
is the new conformal factor, and
\be
\bar{K}_G = -\, 16 \, l^2
\ee
is the {\it conformal} Gaussian curvature. We introduce now the anholonomic tetrad 
field
\be
\bar{h}^a{}_\mu = \bar{\Omega}(x) \, \delta^a{}_\mu.
\label{tetrabar}
\ee
If $\bar{\eta}_{ab}$ denotes the cone spacetime metric, the corresponding de 
Sitter metric can, in this case, be written as
\be
\bar{g}_{\mu \nu} \equiv \bar{h}^a{}_\mu \bar{h}^b{}_\nu \, \bar{\eta}_{ab} =
\bar{\Omega}^2(x) \, \bar{\eta}_{\mu \nu}.
\label{confmet}
\ee
It defines the ``conformal distance'' on de Sitter spacetime, whose quadratic 
interval has the form
\be
d \bar{\tau}^2 \equiv \bar{g}_{\mu \nu} \, dx^\mu dx^\nu =
\bar{\Omega}^2(x) \, \bar{\eta}_{\mu \nu} \,  dx^\mu dx^\nu.
\ee
For $l \to 0$ ($\Lambda \to \infty$),, $dS$ contracts to the cone spacetime $N$, 
and $d \bar{\tau}^2$ reduces to the conformal invariant interval on $N$:
\be
d \bar{\tau}^2 \to d\bar{s}^2 = \bar{\eta}_{\mu \nu} \, dx^\mu dx^\nu.
\ee
On account of the conformal transitivity of the cone spacetime, this is the only 
notion of distance that can be defined on $N$ \cite{cosmo1}.

\subsection{Two family of geodesics}

The Christoffel connection of the de Sitter spacetime metric $g_{\mu \nu}$, given 
by Eq.~(\ref{44}), is
\be
\Gamma^{\lambda}{}_{\mu \nu} = \left[ \delta^{\lambda}{}_{\mu}
\delta^{\sigma}{}_{\nu} + \delta^{\lambda}{}_{\nu}
\delta^{\sigma}{}_{\mu} - \eta_{\mu \nu} \eta^{\lambda \sigma} \right]
\partial_\sigma \left[\ln \Omega(x)\right].
\label{46}
\ee
The corresponding Riemann tensor components are
\be
R^{\mu}{}_{\nu \rho \sigma} = -\, \frac{1}{l^2} \,
\left[\delta^{\mu}{}_{\rho} g_{\nu \sigma} - \delta^{\mu}{}_{\sigma} g_{\nu
\rho} \right].
\label{47}
\ee
As already remarked, there are points in the de Sitter spacetime which cannot be 
connected by any geodesic defined by the metric connection (\ref{46}). The reason 
for this fact is that the metric $g_{\mu \nu}$ defines a ``translational 
distance'' only, whereas the de Sitter spacetime is homogeneous under a 
combination of translation and proper conformal transformations.

On the other hand, the Christoffel connection of the de Sitter spacetime 
$\bar{g}_{\mu \nu}$, given by Eq.~(\ref{confmet}), is
\be
\bar{\Gamma}^{\lambda}{}_{\mu \nu} = \left[ \delta^{\lambda}{}_{\mu}
\delta^{\sigma}{}_{\nu} + \delta^{\lambda}{}_{\nu}
\delta^{\sigma}{}_{\mu} - \bar{\eta}_{\mu \nu} \bar{\eta}^{\lambda \sigma} \right]
\partial_\sigma \left[\ln \bar{\Omega}(x)\right].
\label{46b}
\ee
The corresponding Riemann tensor components are
\be
\bar{R}^{\mu}{}_{\nu \rho \sigma} = - 16 l^2 \,
\left[\delta^{\mu}{}_{\rho} \bar{g}_{\nu \sigma} -
\delta^{\mu}{}_{\sigma} \bar{g}_{\nu \rho} \right].
\label{47b}
\ee
Since the metric $\bar{g}_{\mu \nu}$ defines only a ``conformal distance'', and 
since the de Sitter spacetime is homogeneous under a combination of translation 
and proper conformal transformations, there will again be points on $dS$ which 
cannot be connected by any  of the geodesics defined by the metric connection 
(\ref{46b}). However, the two families of geodesics are complementary in the sense 
that the points that cannot be connected by one family of geodesics can be 
connected by the other family. In other words, any two points of the de Sitter 
spacetime can be connected by a geodesics belonging to one or another of the two 
families of geodesics.

It is important to remark that both Riemann tensors $R^{\mu}{}_{\nu \rho \sigma}$ 
and $\bar{R}^{\mu}{}_{\nu \rho \sigma}$ represent the curvature of the de Sitter 
spacetime. The difference is that, whereas $R^{\mu}{}_{\nu \rho \sigma}$ 
represents the curvature tensor in a parameterization appropriate for  the limit 
of a vanishing cosmological constant, $\bar{R}^{\mu}{}_{\nu \rho \sigma}$ 
represents the curvature tensor in a parameterization suitable for  the limit of 
an infinite cosmological constant. As a straightforward calculation shows, both 
limits yield a spacetime with vanishing curvature: both Minkowski and the cone 
spacetimes are  flat.

\section{de Sitter special relativity}

\subsection{The de Sitter transformations}

The de Sitter transformations can be thought of as rotations in a five-dimensional 
pseudo-Euclidian spacetime,
\be
{\chi'}^C = \Lambda^C{}_D \, \chi^D,
\ee
with $\Lambda^C{}_D$ the group element in the vector representation. Since these 
transformations leave invariant the quadratic form
\be
-\,  \eta_{AB} \chi^A \chi^B = l^2,
\ee
they also leave invariant the length parameter $l$. Their infinitesimal form is
\be
\delta {\chi}^C = \onehalf \, {\mathcal E}^{AB}  L_{AB} \, \chi ^C,
\ee
where ${\mathcal E}^{AB}$ are the parameters and $L_{AB}$ the generators.

\subsubsection{Small cosmological constant}
\label{scc}

For $\Lambda$ small, analogously to the identifications (\ref{dslore}) and 
(\ref{l0}), we define the parameters
\be
\epsilon^{ab} = {\mathcal E}^{ab} \quad \mbox{and} \quad
\epsilon^a = l \, {\mathcal E}^{a4}.
\ee
In this case, in terms of the stereographic coordinates, the infinitesimal de 
Sitter transformation assumes the form
\be
\delta x^c = \onehalf \, \epsilon^{ab} L_{ab} x^c + \epsilon^a \Pi_a x^c,
\ee
or equivalently
\be
\delta x^c = \epsilon^{c}{}_a x^a + \epsilon^a - \frac{\epsilon^b}{4 l^2}
\left(2 x_b x^c - \sigma^2 \delta_b{}^c \right).
\ee
In the limit of a vanishing $\Lambda$, it reduces to the ordinary Poincar\'e 
transformation.

\subsubsection{Large cosmological constant}
\label{lcc}

For $\Lambda$ large, analogously to the identifications (\ref{dSLgbis}) and 
(\ref{linf}), we define the parameters
\be
\bar{\epsilon}^{ab} = \sigma^{4} \, {\mathcal E}^{ab} \quad \mbox{and} \quad
\bar{\epsilon}^{a} = {\mathcal E}^{a4} / 4 l.
\ee
In this case, in terms of the stereographic coordinates, the de Sitter 
transformation assumes the form
\be
\delta x^c = \onehalf \, \bar{\epsilon}^{ab} \bar{L}_{ab} \, x^c + 
\bar{\epsilon}^a \bar{\Pi}_a \, x^c,
\ee
or equivalently
\be
\delta x^c = \bar{\epsilon}^{c}{}_a x^a - \bar{\epsilon^a}
\left(2 x_b x^c - \sigma^2 \delta_b{}^c \right) + 4 l^2 \bar{\epsilon}^a,
\ee
where $ \bar{\epsilon}^{c}{}_a = \bar{\epsilon}^{cb} \, \bar{\eta}_{ba} \equiv 
{\epsilon}^{c}{}_a$. In the limit of an infinite $\Lambda$, it reduces to the a 
conformal Poincar\'e transformation.

\subsection{The Lorentz generators}

Up to now, we have studied the de Sitter transformations on a Minkowski space. In 
what follows we are going to study the form of the corresponding generators on a 
de Sitter spacetime, which is the underlying spacetime of a de Sitter special 
relativity. This will be done by contracting the generators acting in Minkowski 
spacetime with the appropriate tetrads. We begin by considering the Lorentz 
generators.

\subsubsection{Small cosmological constant}

For small $\Lambda$, the generators of an infinitesimal Lorentz transformation are 
(see section \ref{scc})
\be
L_{ab} = \eta_{ac} x^c P_b - \eta_{bc} x^c P_a.
\label{dSLg}
\ee
The corresponding generators acting on a de Sitter spacetime can be obtained by 
contracting $L_{ab}$ with the tetrad $h^a{}_\mu$ given by Eq.~(\ref{tetraze}):
\be
{\mathcal L}_{\mu \nu} \equiv h^a{}_\mu \, h^b{}_\nu \, L_{ab} =
g_{\mu \rho} \, x^\rho \, P_\nu - g_{\nu \rho} \, x^\rho \, P_\mu.
\label{dsgene2}
\ee
These generators are easily found to satisfy the commutation relations
\be
\left[ {\mathcal L}_{\mu \nu}, {\mathcal L}_{\rho \lambda} \right] = g_{\nu \rho} 
{\mathcal L}_{\mu \lambda} + g_{\mu \lambda} {\mathcal L}_{\nu \rho} - g_{\nu 
\lambda} {\mathcal L}_{\mu \rho} - g_{\mu \rho} {\mathcal L}_{\nu \lambda}.
\ee
Even when acting on de Sitter spacetime, therefore, these generators still present 
a well-defined algebraic structure, isomorphic to the usual Lie algebra of the 
Lorentz group. This is a fundamental property in the sense that it allows the 
construction, on the de Sitter spacetime, of an algebraically well defined special 
relativity. This possibility is related to the mentioned fact that, like the 
Minkowski spacetime, the de Sitter spacetime is homogeneous and isotropic 
\cite{jack}.

\subsubsection{Large cosmological constant}

For $\Lambda$ large, the generators of infinitesimal Lorentz transformations are 
(see section \ref{lcc})
\be
\bar{L}_{ab} = \bar{\eta}_{ac} x^c P_b - \bar{\eta}_{bc} x^c P_a.
\label{dSLgbis2}
\ee
On a de Sitter spacetime, their explicit form can be obtained by contracting 
(\ref{dSLgbis2}) with the tetrad $\bar{h}^a{}_\mu$ of Eq.~(\ref{tetrabar}):
\be
\bar{\mathcal L}_{\mu \nu} \equiv \bar{h}^a{}_\mu \, \bar{h}^b{}_\nu \, 
\bar{L}_{ab} =
\bar{g}_{\mu \rho} \, x^\rho \, P_\nu - \bar{g}_{\nu \rho} \, x^\rho \, P_\mu.
\label{dsgene4}
\ee
These generators are easily found to satisfy the commutation relations
\be
\left[ \bar{\mathcal L}_{\mu \nu}, \bar{\mathcal L}_{\rho \lambda} \right] = 
\bar{g}_{\nu \rho} \bar{\mathcal L}_{\mu \lambda} + \bar{g}_{\mu \lambda} 
\bar{\mathcal L}_{\nu \rho} - \bar{g}_{\nu \lambda} \bar{\mathcal L}_{\mu \rho} - 
\bar{g}_{\mu \rho} \bar{\mathcal L}_{\nu \lambda}.
\ee
Like ${\mathcal L}_{\mu \nu}$, therefore, they present a Lorentz-like algebraic 
structure.

\subsection{The de Sitter ``translation'' generators}

Like in the case of the Lorentz generators, the form of the de Sitter 
``translation'' generators acting in the de Sitter spacetime can be obtained by 
contracting $\Pi^a$ and $\bar{\Pi}^a$ with the appropriate tetrad.

\subsubsection{Small cosmological constant}

For $\Lambda$ small, the de Sitter translation generators are given by
\be
\Pi_\mu \equiv h^a{}_\mu \, \Pi^a = \Omega \left[P_\mu - (1/4 l^2) \, K_\mu 
\right],
\ee
where
\be
P_\mu = \partial/\partial x^\mu \quad \mbox{and} \quad K_\mu = \left(
2 \eta_{\mu \rho} \, x^\rho x^\nu - \sigma^2 \delta_\mu{}^\nu \right) P_\nu.
\ee

\subsubsection{Large cosmological constant}

For $\Lambda$ large, on the other hand, they are  
\be
\bar{\Pi}_\mu \equiv \bar{h}^a{}_\mu \, \bar{\Pi}^a = \bar{\Omega} \left[(4 l^2) 
\, P_\mu - K_\mu \right].
\ee
We see from these expressions that the de Sitter spacetime is transitive under a 
combination of the translation and proper conformal generators. For $\Lambda \to 
0$, $\Pi_\mu$ reduce to the usual translation generators of Minkowski spacetime. 
For $\Lambda \to \infty$, $\bar{\Pi}_\mu$ reduce to (minus) the proper conformal 
generators, which define the transitivity on the cone spacetime.

\section{Energy-momentum relations}

Let us consider now the mechanics of structureless  particles on de Sitter 
spacetime. The conserved Noether current associated to a particle of mass $m$ is, 
in this case, the five-dimensional angular momentum \cite{gursey}
\be
\lambda^{A B} = m c \left(\chi^A \; \frac{d \chi^B}{d \tau} - \chi^B \; \frac{d 
\chi^A}{d \tau} \right),
\ee
with $d\tau$ the de Sitter line element (\ref{onod}). In order to make contact 
with the usual definitions of energy and momentum, we rewrite it in terms of the 
stereographic coordinates $\{x^a\}$ and the Minkowski interval $ds$. The result is 
\be
\lambda^{ab} = x^a \, p^b - x^b \, p^a
\label{cc0}
\ee
and
\be
\lambda^{a4} = l p^a - (4 l)^{-1} \, k^a,
\label{cc2}
\ee
where
\be
p^a = m \, c \, \Omega \, \frac{dx^a}{ds} 
\ee
is the momentum, and
\be
k^a = (2 \eta_{cb} \, x^c \, x^a -
\sigma^2 \, \delta_b{}^a) \,  {p}^b
\ee
is the so-called conformal momentum \cite{coleman}. Their form on the de Sitter 
spacetime can be obtained through a contraction with appropriate tetrads.

\subsection{Small cosmological constant}

For $\Lambda \, l_P^2 \to 0$, analogously to the generators, we define the de 
Sitter momentum
\be
\pi^a \equiv \frac{\lambda^{a4}}{l} =
{p}^a - \frac{{k}^a}{4 l^2} \,\, .
\label{dstra8}
\ee
The corresponding spacetime version is
\be
\pi^\mu \equiv h_a{}^\mu \, \pi^a =
{p}^\mu - \frac{{k}^\mu}{4 l^2},
\label{dstra4}
\ee
where
\be
p^\mu = m \, c \, \frac{dx^\mu}{ds} 
\ee
is the Poincar\'e momentum, and 
\be
k^\mu = (2 \eta_{\lambda \rho} \, x^\rho \, x^\mu -
\sigma^2 \, \delta_\lambda{}^\mu) \,  {p}^\lambda
\ee
is the corresponding conformal Poincar\'e momentum. We remark that $\pi^\mu$ is 
the conserved Noether momentum related to the transformations generated by 
$\Pi_a$. Similarly to the identification $p^\mu = T^{\mu 0}$, with $T^{\mu \nu}$ 
the energy-momentum current, the conformal momentum $k^\mu$ is defined by $k^\mu = 
K^{\mu 0}$, with $K^{\mu \nu}$ the conformal current \cite{coleman}. 

If we define the de Sitter mass $M$ as the first Casimir invariant of the de 
Sitter group \cite{gursey}
\be
g_{\mu \nu} \pi^\mu \pi^\nu \equiv M^2 \, c^2,
\label{dsdr0}
\ee
and write the components of $\pi^\mu$ as $(i, j, \dots = 1, 2, 3)$
\be
\pi^\mu \equiv \left(\frac{E}{c}, P^i \right),
\ee
with $E$ and $P^i$ the de Sitter notions of energy and momentum, the kinematic 
relation in the de Sitter relativity can be written in the form
\be
\frac{E^2}{c^2} = P^2 = M^2 \, c^2.
\label{dsdr}
\ee

However, in order to make contact with ordinary special relativity, it is 
convenient to rewrite the dispersion relation (\ref{dsdr}) in terms of the usual 
notions of mass, energy and momentum. To do it, we observe first that, whereas
\be
\pi^0 \equiv {p}^0 - \frac{{k}^0}{4 l^2}
\ee
represents the energy, the space components
\be
\pi^i \equiv {p}^i - \frac{{k}^i}{4 l^2}
\ee
represent the momentum. The presence of a cosmological constant, therefore, 
changes the usual definitions of energy and momentum \cite{aap,hossen}. When 
written in terms of these usual notions, the energy-momentum relation 
(\ref{dsdr0}) becomes
\be
g_{\mu \nu} \pi^\mu \pi^\nu = \Omega^2 \, \eta_{\mu \nu} \left(
p^\mu p^\nu - \frac{1}{2 l^2} p^\mu k^\nu + \frac{1}{16 l^4} k^\mu k^\nu \right).
\label{emr1}
\ee
The components of the Poincar\'e momentum $p^\mu$ are
\be
p^\mu = \left(\frac{\varepsilon_p}{c} ,  p^i \right),
\ee
where $\varepsilon_p$ and $p^i$ are the usual Poincar\'e energy and momentum, 
respectively, related by  $\eta_{\mu \nu} \, p^\mu p^\nu = m^2 c^2$, where $m^2 
c^2$ is the first Casimir invariant of the Poincar\'e group. Analogously, the 
components of the conformal momentum $k^\mu$ can be written in the form
\be
k^\mu = \left(\frac{\varepsilon_k}{c} , k^i \right),
\ee
with $\varepsilon_k$ the conformal notion of energy, and $k^i$ the space 
components of the conformal momentum. The conformal momentum satisfies $\eta_{\mu 
\nu} \, k^\mu k^\nu = \bar{m}^2 c^2$, where $\bar{m}^2 c^2$ is the first Casimir 
invariant of the conformal Poincar\'e group, with $\bar{m}^2 = \sigma^{4} m^2$ the 
conformal equivalent of the mass \cite{dssr}. Using the expressions above, the 
relation (\ref{emr1}) becomes
\be
\frac{\varepsilon_p^2}{c^2} - {p}^{2} = m^2  c^2 + \frac{1}{2 l^2} \left[
\frac{\varepsilon_p \varepsilon_k}{c^2} - \vec{p} \cdot \vec{k} - m \bar{m} c^2 -
\frac{1}{8 l^2} \left(\frac{\varepsilon_k^2}{c^2} - k^2 - \bar{m}^2 c^2 \right) 
\right].
\label{hedr}
\ee
For small values of $\Lambda$, the de Sitter length parameter $l$ is large, and 
the modifications in the energy-momentum relation will be small. Up to first order 
in $\Lambda$, we get
\be
\frac{\varepsilon_p^2}{c^2} - {p}^2 \simeq m^2 \, c^2 +
\frac{1}{2 l^2} \left[\frac{\varepsilon_p \varepsilon_k}{c^2} - \vec{p} \cdot 
\vec{k} - m \, \bar{m} \, c^2 \right].
\ee
In the limit of a vanishing cosmological constant, the ordinary notions of energy 
and momentum are recovered, and the de Sitter relativity reduces to the ordinary 
special relativity, in which the Poincar\'e symmetry is exact. The energy-momentum 
relation, in this case, reduces to the usual expression
\be
\frac{\varepsilon_p^2}{c^2} - {p}^2 = m^2 \, c^2.
\label{poincaredr}
\ee

\subsection{Large cosmological constant}

For $\Lambda \, l_P^2 \to 1$, analogously to the generators, we define the de 
Sitter momentum
\be
\pi^a \equiv 4 l \, {\lambda^{a4}} =
4 l^2 {p}^a - {k}^a.
\label{dstra9}
\ee
The corresponding spacetime version is
\be
\bar{\pi}^\mu \equiv \bar{h}_a{}^\mu \, \bar{\pi}^a =
\frac{4 l^2}{\sigma^2} \left( 4 l^2 {p}^\mu - {k}^\mu \right).
\label{dstra10}
\ee
We remark that $\bar{\pi}^\mu$ is the conserved Noether momentum related to the 
transformations generated by $\bar{\Pi}_a$. Defining the de Sitter conformal mass 
$\bar{M}$ by
\be
\bar{g}_{\mu \nu} \bar{\pi}^\mu \bar{\pi}^\nu \equiv \bar{M} \, c^2,
\label{condsdr0}
\ee
and writing the components of $\bar{\pi}^\mu$ as
\be
\bar{\pi}^\mu \equiv \left( \frac{\bar{E}}{c}, \bar{K}^i \right),
\ee
with $\bar{E}$ and $\bar{K}^i$ the de Sitter notions of conformal energy and 
conformal momentum, the kinematic relation in this case assumes the form
\be
\frac{\bar{E}^2}{c^2} = \bar{K}^2 + \bar{M}^2 \, c^2.
\label{dsdrcon}
\ee

Like in the previous case, it is convenient to rewrite the dispersion relation 
(\ref{dsdrcon}) ini terms of the usual notions of conformal mass, energy and 
momentum. To do it, we observe that, whereas
\be
\bar{\pi}^0 = \frac{4 l^2}{\sigma^2} \, ( 4 l^2 {p}^0 - {k}^0)
\ee
represents the conformal energy, the space components
\be
\bar{\pi}^i = \frac{4 l^2}{\sigma^2} \, ( 4 l^2 {p}^i - {k}^i)
\ee
represent the conformal momentum. In this case, the energy-momentum relation 
(\ref{condsdr0}) becomes
\be
\bar{g}_{\mu \nu} \bar{\pi}^\mu \bar{\pi}^\nu = 16 l^4 \, \bar{\Omega}^2 \, 
\sigma^{-8} \, \eta_{\mu \nu} \left[16 l^4 p^\mu p^\nu - 8 l^2 p^\mu k^\nu + k^\mu 
k^\nu \right],
\label{emr5}
\ee
or equivalently,
\be
\frac{\varepsilon_k^2}{c^2} - {k}^2 = \bar{m}^2 c^2 + 8 l^2 \left[
\frac{\varepsilon_p \varepsilon_k}{c^2} - \vec{p} \cdot \vec{k} - m \bar{m} c^2 -
2 l^2 \left(\frac{\varepsilon_p^2}{c^2} - p^2 - m^2 c^2 \right) \right].
\label{ledr}
\ee
For large values of the cosmological constant, the de Sitter length parameter $l$ 
is small. Up to first order in $l^2$, we get
\be
\frac{\varepsilon_k^2}{c^2} - {k}^2 \simeq \bar{m}^2 \, c^2 + 8 \, l^2 \left[
\frac{\varepsilon_p \varepsilon_k}{c^2} - \vec{p} \cdot \vec{k} - m \, \bar{m} \, 
c^2 \right].
\ee
In the formal limit $\Lambda \, l_P^2 \to \infty$, only the conformal notions of 
energy and momentum will remain, and de Sitter relativity will reduce to the pure 
conformal special relativity. In this case, the energy-momentum relation acquires 
the conformal special relativistic form
\be
\frac{\varepsilon_k^2}{c^2} - {k}^2 = \bar{m}^2 \, c^2.
\ee

\section{Final remarks}

A non-vanishing cosmological term introduces the conformal generators in the 
definition of spacetime transitivity. As a consequence, the conformal 
transformations will naturally be incorporated in the kinematics of spacetime, and 
the corresponding conformal current will appear as part of the Noether conserved 
current. Of course, for a small enough cosmological term, the conformal 
modifications become negligible and ordinary physics remains valid. For large 
values of $\Lambda$, however, the conformal contributions to the physical 
magnitudes cannot be neglected, and these contributions will give rise to deep 
conceptual changes. For example, ordinary special relativity, which is based on 
the Poincar\'e group, will no longer be true, and must be replaced by a new 
special relativity based on the de Sitter group. The physical tangent space at 
each point of any spacetime will consequently be converted into an osculating de 
Sitter spacetime.

Due to the fact that the de Sitter spacetime is transitive under a combination of 
translation and proper conformal transformations, the de Sitter special relativity 
can be viewed as made up of two different relativities: the usual one, related to 
translations, and a conformal one, related to proper conformal transformations. It 
is a single relativity interpolating these two extreme limiting cases. For small 
values of $\Lambda$, ordinary special relativity will prevail over the conformal 
one, and the Poincar\'e symmetry will be weakly deformed. In the contraction limit 
of a vanishing  $\Lambda$,  de Sitter relativity reduces to the ordinary special 
relativity, and the underlying spacetime reduces to the Minkowski space $M$, which 
is transitive under translations. For large values of $\Lambda$, on the other 
hand, conformal relativity will prevail over the usual one, and the Poincar\'e 
symmetry will be strongly deformed. In the contraction limit of an infinite 
$\Lambda$, de Sitter special relativity reduces to a conformal relativity, and the 
underlying spacetime reduces to a maximally symmetric cone spacetime, which is 
transitive under proper conformal transformations only. Although translations, and 
consequently the usual notion of distance, cannot be defined on such spacetime, it 
can be said to be conformally infinite. Physically, the cone spacetime represents 
a universe in which all energy is in the form of dark energy \cite{confor}.

\subsection{Relation to doubly special relativity}

The de Sitter group can be interpreted as a particular deformation of the 
Poincar\'e group. It is related to Poincar\'e through the contraction limit of a 
vanishing cosmological constant, in the very same way the Galilei group is related 
to the Poincar\'e group through the contraction limit of an infinite velocity of 
light. A special relativity based on the de Sitter group, therefore, gives rise to 
a specific kind of deformed (or doubly, as it has been called) special relativity 
(DSR) \cite{dsr}. In the usual approach to DSR, the Poincar\'e symmetry is 
deformed through the agency of a dimensional parameter $\kappa$, proportional to 
the Planck length. Such a deformation implies that, in the high energy limit, the 
new ensuing relativity principle requires that physics become invariant, not under 
Poincar\'e, but under a ``$\kappa$-deformed'' Poincar\'e group, which reduces to 
the standard Poincar\'e group in the low energy limit.

A crucial difference between the standard formulations of DSR and a de Sitter 
special relativity concerns the structure of the symmetry groups. In a $\kappa$-
deformed Poincar\'e group, the Lorentz subgroup is deformed by the $\kappa$ 
parameter, and consequently the corresponding relativity will include a violation 
of the Lorentz symmetry at the scales where the theory is supposed to hold up. On 
the other hand, in a de Sitter special relativity, the Lorentz subgroup remains 
untouched, and consequently the Lorentz symmetry remains a true symmetry at all 
scales. Instead of Lorentz, the presence of a non-vanishing $\Lambda$ produces a 
violation of the translation symmetry. In fact, when Minkowski spacetime is 
replaced by de Sitter, the corresponding kinematic groups change from Poincar\'e 
to de Sitter. Since, algebraically speaking, the only difference between these two 
groups is the replacement of $P_a$ by a linear combination of $P_a$ and $K_a$, the 
net result of this change is ultimately the {\it breakdown of translational 
invariance}. From the experimental point of view, therefore, a de Sitter special 
relativity could be probed by looking for possible violations of translational 
invariance. This could be done, for example, by applying the same techniques used 
in the search for possible violations of Lorentz and CPT symmetries in high energy 
processes \cite{lcpt}. For small values of $\Lambda$, as we have seen, the 
homogeneity of spacetime is preponderantly defined by the translation generators, 
which means that the violation of the translation invariance will be very small. 
Only when $\Lambda$ is large this violation is expected to be relevant.

Another important difference is that, in the standard $\kappa$-deformed 
formulations of doubly special relativity, although energy and momentum keep their 
ordinary special relativistic notions, they are assumed to satisfy a deformed 
dispersion relation. As a consequence, a consistent notion of total energy and 
momentum, as well as a conservation law for them, is lacking in these theories 
\cite{kgn}. In a de Sitter special relativity, on the other hand, a modified but 
precise notion of momentum and energy is provided: they are the Noether currents 
associated to the de Sitter symmetry. There is a clear relation between the 
symmetry generators (\ref{cp0}-\ref{dstra}) and the conserved currents (\ref{cc0}-
\ref{cc2}). The resulting deformed dispersion relations for the particle's energy 
and momentum, given by Eqs.~(\ref{hedr}) and (\ref{ledr}), are consequently 
relations between conserved quantities. Since the de Sitter current is a linear 
combination of the ordinary momentum $p^\mu$ and the conformal momentum $k^\mu$, 
the dispersion relations turn out to depend on these two four-vectors. Notice, 
however, that neither $p^\mu$ nor $k^\mu$ is conserved: only the de Sitter 
momentum is conserved.

It is also worth mentioning that, in the very same way as it happens with the 
ordinary special relativity dispersion relation (\ref{poincaredr}), the de Sitter 
dispersion relations (\ref{hedr}) and (\ref{ledr}) are invariant under a 
simultaneous re-scaling of mass, energy and momentum. On the other hand, because 
it includes non-quadratic terms in the momentum, the dispersion relations of the 
usual formulations of DSR are not invariant under such a re-scaling. As a 
consequence, they present the so called ``soccer-ball problem'' \cite{soccer}. 
This is not the case of the dispersion relations of the de Sitter special 
relativity, whose scale invariance make them true for elementary particles (as it 
is argued since they must be relevant for discussing elementary particle 
processes), as well as for macroscopic objects like, for example, a soccer-ball. 

\subsection{Causal structure and quantum gravity}

If our current theories of particle physics based on spontaneously broken symmetry 
and phase transition are correct, there must have been some periods in the history 
of the Universe in which the value of $\Lambda$, and hence of the scale energy 
$E_{\Lambda}$, were large. For example, in the electroweak epoch characterized by 
$\Lambda_{EW}$, the kinematics of a typical electroweak process with energy 
$E_{\Lambda_{EW}}$, according to the de Sitter special relativity, must have been 
strongly influenced by $\Lambda_{EW}$. In fact, as we have seen, a large 
cosmological constant would produce significant changes in the definitions of 
energy and momentum, as well as in the kinematic relations satisfied by them. 
These changes could modify significantly the physics that should be applied in the 
study the early universe. To understand these changes, it is important to notice 
that the conformal symmetry of high-$\Lambda$ spacetimes is not just a symmetry 
presented by some physical systems, like for example the conformal symmetry of the 
electromagnetic field, but a {\it kinematic symmetry}, that  is, a symmetry 
related to the relativity principle governing the physics at that scale.  

On the other hand, if we take spontaneously broken symmetries as the primary 
source of a non-vanishing $\Lambda$, it is conceivable that a high energy 
experiment could modify the {\it local structure of space-time for a short period 
of time}, in such a way that the immediate neighborhood of a high energy collision 
would depart from the Minkowski space and become a de Sitter spacetime. According 
to this scheme, there would be a connection between the energy scale of the 
experiment and the local value of $\Lambda$ \cite{mansouri}. For energies of the 
order of 200~GeV, corresponding to the electroweak phase transition, the de Sitter 
parameter is $l_{\Lambda_{EW}} \sim 0.25$ cm, which is equivalent to 
$E_{\Lambda_{EW}}\sim 10^{-4}$~eV. For high energy experiments of order 20~TeV, 
one finds $E_{\Lambda_{TeV}} \sim 1$~eV. And, for energies of order 1000~TeV, we 
have $E_{\Lambda}\sim 2500$~eV. For particles of small mass, such as neutrinos, 
there would be significant changes in the kinematics at very high energies, which 
could eventually be tested in the existing or planned colliders \cite{mansouri}.

Another important point refers to some possible implications to quantum gravity. 
Due to the homogeneous character of the de Sitter spacetime, the Lorentz 
generators in this spacetime still present a well defined algebraic structure, 
isomorphic to the usual Lie algebra of the Minkowski Lorentz group. This means 
that the Lorentz symmetry remains a sub-symmetry in a de Sitter relativity, and 
consequently the velocity of light $c$ is left unchanged by a de Sitter 
transformation. Since it also leaves unchanged the length parameter $l$, a de 
Sitter transformation leaves unchanged both $c$ and $l$. This property has 
important consequences for causality. As is well known, the constancy of $c$ 
introduces a causal structure in spacetime, defined by the light cones. The 
presence of the de Sitter length parameter $l$ adds to that structure some further 
restrictions on the causal structure of spacetime. To see that, let us remember 
that the de Sitter spacetime has a horizon, which restricts the causal region of 
each observer. In terms of the stereographic coordinates, this horizon is 
identified by
\be
x^2 + y^2 + z^2 = l^2/\Omega^2 \quad \mbox{and}
\quad (x^0)^2 = l^2 (2 - 1/\Omega)^2.
\ee
For small $\Lambda$, the horizon tends to infinity, and there are no significant 
causal changes. For large values of $\Lambda$, however, the causal domain of each 
observer --- restricted by the horizon --- becomes small. Considering again a 
possible $\Lambda$-dependence of high energy processes, for experiments of the 
order of the Planck energy, this region would be of the order of the Planck length 
$l_P$. At this scale, therefore, the large value of $\Lambda$ would introduce deep 
changes in the causal structure of spacetime \cite{hbar}. In particular, a de 
Sitter transformation, at this scale, would leave both $c$ and $l_P$ unchanged. 
Finally, it is interesting to observe that, since the area $A_{dS}$ of the de 
Sitter horizon is proportional to $l^2$, the entropy associated to this surface 
will be proportional to the logarithm of the number of states
\[
n = A_{dS} / l_P^2 \sim {l^2}/{l_P^2}.
\]
At the Planck scale, $n \sim 1$, which yields a vanishing entropy for the de 
Sitter horizon. This property provides a contact between de Sitter special 
relativity and quantum gravity \cite{nzcc}, and could be important for the 
understanding of the spacetime structure at the Planck scale.

\begin{theacknowledgments}
The authors would like to thank M. Novello for the invitation to present this 
lecture at the School. They would also like to thank FAPESP, CNPq and CAPES for 
financial support.
\end{theacknowledgments}


\end{document}